\documentclass[aps, 10pt, notitlepage, twocolumn, superscriptaddress,nofootinbib,longbibliography]{revtex4-2}

\usepackage{dcolumn}
\usepackage{bm}
\usepackage{ifpdf}
\usepackage{hyperref}
\usepackage{float}
\usepackage{bm}
\usepackage{xcolor,color,graphicx,graphics}
\usepackage[spanish,english]{babel}
\usepackage[latin1]{inputenc}
\usepackage[OT1]{fontenc}
\usepackage{latexsym,amssymb,amsmath,amsfonts}
\usepackage{makeidx}
\usepackage{epsfig,subfigure}
\usepackage{epstopdf}
\usepackage{mathrsfs}
\hypersetup{colorlinks=true, linkcolor=blue, citecolor=green}
\usepackage{enumerate}
\usepackage{xcolor}
 \usepackage{multirow}

\definecolor{red}{rgb}{1,0,0}

\def\+{^\dagger}

\def\<{\leftarrow}
\def\>{\rightarrow}

\def\({\left(}
\def\){\right)}

\newcommand{\bi}{\begin{itemize}} 				\newcommand{\ei}{\end{itemize}}
\newcommand{\benu}{\begin{enumerate}} 		\newcommand{\enu}{\end{enumerate}}
\newcommand{\bd}{\begin{dinglist}{0}}     \newcommand{\ed}{\end{dinglist}}
\newcommand{\bfig}{\begin{figure}[htbp]}  \newcommand{\efig}{\end{figure}}
        			
\newcommand{\bc}{\begin{center}} 				  \newcommand{\ec}{\end{center}}
\newcommand{\be}{\begin{equation}} 				\newcommand{\ee}{\end{equation}}
\newcommand{\bsub}{\begin{subequations}}  \newcommand{\esub}{\end{subequations}}
\newcommand{\ben}{\begin{eqnarray}} 			\newcommand{\een}{\end{eqnarray}}
\newcommand{\ba}[1]{\begin{array}{#1}} 		\newcommand{\ea}{\end{array}}
\newcommand{\bea}{\begin{equation}\begin{array}{rcl}}
\newcommand{\eea}{\end{array}\end{equation}}

\begin{document}

\color{black}       

\title{Stellar structure in $f(R,T)$ gravity: some exact solutions}

\author{Aliya Batool }
\email[Email: ]{syedaaliyabatool123@gmail.com}
\affiliation{Department of Mathematics,
University of Okara, Okara-56300 Pakistan} 
\author{Abdul Malik Sultan }
\email[Email: ]{ams@uo.edu.pk, maliksultan23@gmail.com}
\affiliation{Department of Mathematics,
University of Okara, Okara-56300 Pakistan}
\author{Gonzalo J. Olmo}
\email[Email: ]{gonzalo.olmo@uv.es}
\affiliation{Departamento de F\'isica Te\'orica and IFIC, Centro Mixto Universidad de Valencia - CSIC. Universidad de Valencia, Burjassot-46100, Valencia, Spain}
\affiliation{Universidade Federal do Cear\'a (UFC), Departamento de F\'isica,
Campus do Pici, Fortaleza - CE, C.P. 6030, 60455-760 - Brazil}

\author{Diego Rubiera-Garcia} 
\email[Email: ]{ drubiera@ucm.es}
\affiliation{Departamento de F\'isica Te\'orica and IPARCOS, Universidad Complutense de Madrid, E-28040 Madrid, Spain}

\begin{abstract}
We find some exact solutions for constant-density and quark matter equations of state in stellar structure models framed within the $f(R,T)=R+\lambda \kappa^2 T$ theory of gravity, where $R$ is the curvature scalar, $T$ the trace of the stress-energy tensor, and $\lambda$ some constant. These solutions correspond to specific values of the constant $\lambda$, and represent different compactness states of the corresponding stars, though only those made of quark matter can be regarded as physical. The latter modify the compactness (Buchdahl) limit of neutron stars upwards for $\lambda>0$ until saturating the one of black holes. Our results show that it is possible to find useful insights on stellar structure in this class of theories, a fact that could be used for obtaining constraints on limiting masses such as the minimum hydrogen burning mass.
\end{abstract}

\maketitle

\section{Introduction}  \label{sec1}

The study of modified theories of gravity beyond General Relativity (GR) has blossomed in the last two decades driven by astrophysical \cite{Berti:2015itd} and cosmological observations \cite{Bull:2015stt} that challenge the standard model of a universe governed by the dynamics of GR under the influence of the matter sources found in laboratory. The need for new matter and energy sources with unusual properties to fit observational data could be minimized or even avoided if the effective large scale dynamics could differ from that of GR, offering in this way a completely different interpretation of the data. In this debate, many different gravity models have been considered, from theories in which gravity is mediated by a metric tensor and other fundamental fields, like scalars \cite{Sotiriou:2006hs,Gleyzes:2014dya,Langlois:2017dyl} and/or vector fields \cite{Kimura:2016rzw,Heisenberg:2018acv}, to alternative geometric frameworks of the metric-affine type \cite{Olmo:2011uz}, of Finsler spaces \cite{Pfeifer:2011xi}, noncommutative geometries \cite{Connes:1996gi}, and so on (see \cite{DeFelice:2010aj,Nojiri:2017ncd} for some reviews). 

Among the geometric extensions of GR, theories with non-minimal matter-curvature couplings have also been widely investigated (see \cite{LoboBook} for a full account of this topic). The generic violation of stress-energy conservation that these theories bring with them can generate new forces that could have an impact in cosmological and astrophysical scenarios of interest. The case of $f(R,T)$ theories (where $R$ stands for the curvature scalar and $T$ for the trace of the stress-energy tensor)  \cite{Harko:2011kv} is particularly interesting in this regard because they allow the construction of minimal extensions of GR in which the very presence of matter may induce non-trivial modifications of the field equations with effects at all scales, including laboratory experiments, cosmological scenarios \cite{Jamil:2011ptc,Alvarenga:2013syu,Zaregonbadi:2016xna} solar system dynamics \cite{Shabani:2014xvi}, and even in stellar structure models \cite{Das:2017rhi,Bhatti:2020rzr,Rej:2021ngp,Tangphati:2022mur}. Regarding the latter, $f(R,T)$ theories of gravity lead to a modified system of Tolman-Oppenheimer-Volkov (TOV) equations of hydrostatic equilibrium, thus providing a different set of predictions for important observational features such as the mass-radius relation, as well as the compactness (Buchdahl) limit, which sets the maximum amount of mass within a given radius that a fluid-made rotating star can attain \cite{Olmo:2019flu}. Other  limits, such as the minimum mass required for stable hydrogen burning in stars \cite{Sakstein:2015zoa,Sakstein:2015aac,Olmo:2019qsj,Olmo:2021yul} can also be affected by the $f(R,T)$ dynamics.

The main aim of the present work is to consider a minimal $f(R,T)$ model of the form $f(R,T)=R+\lambda \kappa^2 T$ (where  $\kappa^2=8\pi G$ represents Newton's constant in suitable units) and $\lambda$ some constant,  and explore its impact in the properties of compact stellar objects. Studies of this type have already been considered in \cite{Pretel:2020oae,Pretel:2022dbx,Tangphati:2022mur} where numerical methods were used to investigate the properties of pulsating stars and charged stars with quark matter. Our interest here is in determining the limits of compactness in these stars, for which we focus on the possibility of obtaining exact analytical solutions. As we will see, unlike in GR, were solutions of the TOV equations can be found analytically and used to establish the so-called Buchdahl's limit on the compactness of stars with monotonic density profile with a maximum at the center, in $f(R,T)$ theories it is non-trivial to find analytical solutions, though some can be found under certain restrictions  \cite{Kumar:2021vqa}. We focus on two such solutions: one for constant-density stars for a specific value of the parameter $\lambda$, and another corresponding to an equation of state for quark matter, for which all values of $\lambda$ can be considered. 

The contents of the paper are organized as follows: the field equations and stellar structure equations are presented in Sec.\ref{sec:FEqs}, followed by a derivation of constant density stars in Sec.\ref{sec:Constant}, and of quark matter stars in Sec.\ref{sec:Quark}. The paper is closed in Sec.\ref{sec:Final}, where some comments and conclusions are provided.


\section{Field equations}\label{sec:FEqs}

The field equations for $f(R,T)$ theories take the form \cite{Harko:2011kv} 
\begin{equation}
f_R R_{\mu\nu}-\frac{f}{2}g_{\mu\nu}+(g_{\mu\nu} \Box- \nabla_\mu \nabla_\nu)f_R= (\kappa^2-f_T)T_{\mu\nu}-f_T\Theta_{\mu\nu} \ ,
\end{equation}
where we are using the notation $f_R \equiv df/dR$ and $f_T \equiv df/dT$. Furthermore, we are assuming as the matter source a perfect fluid with the usual stress-energy tensor
\begin{equation}
T_{\mu\nu}=(\rho+P)u_\mu u_\nu-P g_{\mu\nu} \ ,
\end{equation}
with $\rho$ the energy density and $P$ the pressure of the fluid, and we are assuming the normalization $u^\mu u_\mu=+1$ for the time-like vector $u^\mu$. For this choice of fluid, then we have the object 
\begin{equation}
\Theta_{\mu\nu}=-2T_{\mu\nu}-P g_{\mu\nu} \ ,
\end{equation}

Particularizing the above equations to the family of theories $f(R,T)=R+\lambda\kappa^2 T$, they can be written simply as
\begin{equation} \label{eq:eom}
G_{\mu\nu} \equiv R_{\mu\nu}-\frac{R}{2}g_{\mu\nu}=(1+\lambda)\kappa^2 T_{\mu\nu}+\lambda\kappa^2\left(\frac{\rho-P}{2}\right) \ ,
\end{equation}
which represent a set of modified Einstein equations with additional matter terms on the right-hand side. Interestingly, the vacuum field equations retrieve the same solutions as those of GR and, in particular, the Schwarzschild solution can be regarded as the most general spherically symmetric vacuum solution within this setting.  

On the other hand, it is evident that the last term of the field equations (\ref{eq:eom})  will, in general, lead to violations of energy conservation, which implies the existence of extra forces in the fluid equations of motion (see also \cite{PerezTeruel:2022wsu} for other theories with this property). In fact, the conservation of the Einstein tensor, $\nabla^{\mu}G_ {\mu\nu}=0$, implies that the stress-energy tensor in this theory must satisfy
\begin{equation}\label{eq:DTmn}
\nabla_\mu {T^\mu}_\nu=-\frac{\lambda }{2(1+\lambda)}\nabla_\nu\left(\rho-P\right)  \ ,
\end{equation}
which is perhaps the most interesting novel feature of this class of theories. We will now explore how this affects the TOV stellar structure equations. 

\subsection{Modified TOV and structure equations}

Let us consider a spherically symmetric line element of the form 
\begin{equation}
ds^2=A(r)e^{2\Phi(r)}dt^2-\frac{1}{A(r)}dr^2-r^2d\Omega^2 \ ,
\end{equation}
where $A(r)$ and $\Phi(r)$ are functions of the radial coordinate only and $d\Omega^2=d\theta^2 + \sin^2 \theta d\varphi^2$ is the line element on the two-spheres. With this line element Eq.(\ref{eq:DTmn}) leads, after a little bit of algebra, to the modified TOV equation
\begin{equation}
P_r+\frac{(\rho+P)}{2}\left(\frac{A_r}{A}+2\Phi_r\right)=\frac{\lambda}{2(1+\lambda)}\left(\rho_r-P_r\right) \ .
\end{equation}
where $\rho_r \equiv d\rho/dr$, $P_r \equiv dP/dr$, and similarly for the metric functions. Assuming now an equation of state of the form $\rho=\rho(P)$, 
we can rewrite the above equation as:
\begin{equation}\label{eq:TOV}
\left(1+\frac{\lambda\left(1-\rho_P\right)}{2(1+\lambda)}\right)\frac{P_r}{(\rho+P)}=-\frac{1}{2}\left(\frac{A_r}{A}+2\Phi_r\right) \ ,
\end{equation}
where $\rho_P\equiv d\rho/dP$. As expected, when $\lambda=0$ one recovers the usual TOV equation of GR for a perfect fluid.  It is easy to see that the right-hand side of this equation is the total radial derivative of the function $-\log (Ae^{2\Phi})/2$, while the left-hand side can be integrated in exact analytical form for specific forms of the function $\rho(P)$. For instance, if one takes the choice
\begin{equation}\label{eq:EOS}
\rho=\rho_0+\sigma P^\alpha \ ,
\end{equation}
where $\rho_0,\sigma$ and $\alpha$ are constants, for some values of $\alpha$ one can find explicit solutions. Taking, for simplicity,  the case $\alpha=1$, which represents an equation of state for quark matter (such as in the MIT bag model \cite{Chodos:1974je}), we find that Eq.(\ref{eq:TOV}) can be integrated as
\begin{equation}\label{eq:quarkEOS0}
Ae^{2\Phi}=\frac{C}{\left(\rho_0+(1+\sigma)P\right)^\gamma} \ ,
\end{equation}
where we have defined the constant
\begin{equation} \label{eq:gamma}
\gamma=\frac{2}{1+\sigma}\left(1+\frac{\lambda(1-\sigma)}{2(1+\lambda)}\right) \ ,
\end{equation}
gathering all the constants of the problem besides an integration constant $C$ which can be determined by imposing agreement with the Schwarzschild solution at the surface of the star, namely, $A_\odot\equiv A(r=R_\odot)=1-2M_\odot/R_\odot$ and $\Phi(r=R_\odot)=0$ when $P(r=R_\odot)=0$, where $R_{\odot}$ denotes the stellar's surface radius.. This means that $C=A_\odot\rho_0^\gamma$.

Turning back to the field equations (\ref{eq:eom}), considering the ${G^t}_t$ and ${G^r}_r$ components, and introducing the usual mass ansatz $A(r)=1-2M(r)/r$, elementary manipulations lead to the pair of equations
\begin{eqnarray}
\frac{2A \Phi_r}{r}&=& \kappa^2(1+\lambda)\left(\rho+P\right) \label{eq:Phir0} \  \\
\frac{2M_r}{r^2}&=& \kappa^2\left(1+\frac{3\lambda}{2}\right)\rho-\frac{\lambda \kappa^2 P }{2}  \label{eq:Mr0} \ .
\end{eqnarray}
We can now use Eqs.(\ref{eq:Phir0}), (\ref{eq:Mr0}), and (\ref{eq:TOV}) together with some equation of state $\rho=\rho(P)$ to find solutions for stellar objects. In \cite{Moraes:2015uxq} numerical methods were used to study mass-radius relations and the maximum mass   in this theory. Here we will consider the particular case (\ref{eq:EOS}) to obtain two exact analytical solutions, which will allow us to extract some useful information to compare the predictions of this family of $f(R,T)$ theories with GR. 

\section{Constant density stars}\label{sec:Constant}

If we set $\lambda=0$ in the equations of the previous section and take a constant density configuration, $\rho=\rho_0$ for $r\leq R_\odot$ and $\rho=0$ for $r>R_\odot$, one can use Eq.(\ref{eq:Mr0}) to obtain
\begin{equation}
M(r)=\frac{M_\odot}{R_\odot^3} r^3
\end{equation}
with $M_\odot/R_\odot^3\equiv \kappa^2\rho_0/6$. To solve for $\Phi(r)$, we insert this function $M(r)$ into $A(r)$ and use Eq.(\ref{eq:quarkEOS0}) (with $\lambda=0$ and $\sigma=0$) to rewrite $\rho+P$ as 
\begin{equation}\label{eq:PressGR}
\rho_0+P=\frac{\rho_0 A^{1/2}_\odot}{A^{1/2}(r)e^\Phi} \ ,
\end{equation}
where we are denoting $A(R_\odot)\equiv A_\odot$. Plugging this in Eq.(\ref{eq:Phir}), we find
\begin{equation}
e^\Phi \Phi_r =\frac{3M_\odot A^{1/2}_\odot}{R_\odot^3}\frac{r}{A^{3/2}} \ ,
\end{equation}
which can be integrated yielding
\begin{equation}
e^{\Phi(r)}=\left\{\begin{array}{ll}
\frac{3A^{1/2}_\odot-A^{1/2}(r)}{2A^{1/2}(r)}  & r\leq R_\odot \\ 
\hspace{1cm}\\
1 & r\geq R_\odot \end{array}\right. \ .
\end{equation}
The line element inside this constant density star becomes
\begin{equation}
ds^2=\frac{\left(3A^{1/2}_\odot-A^{1/2}(r)\right)^2}{4}dt^2-\frac{1}{A(r)}dr^2-r^2d\Omega^2 \ ,
\end{equation}
where
\begin{equation}\label{eq:AGR}
A(r)=\left\{\begin{array}{ll}
1-\frac{2M_\odot r^2}{R_\odot^3} & r\leq R_\odot \\ 
\hspace{1cm}\\
1-\frac{2M_\odot}{r} & r> R_\odot \end{array}\right. \ .
\end{equation}
The pressure profile can be obtained from Eq.(\ref{eq:PressGR}) and takes the form
\begin{equation}
P(r)=\rho_0\frac{A^{1/2}(r)-A^{1/2}_\odot}{3A^{1/2}_\odot-A^{1/2}(r)} \ .
\end{equation}
The pressure at the center is given by 
\begin{equation}
P_c=\rho_0\frac{(1-A^{1/2}_\odot)}{(3A^{1/2}_\odot-1)} \ ,
\end{equation}
where we used $A^{1/2}(0)=1$. This pressure diverges in the limit $M_\odot/R_\odot \to 4/9$, which is known as Buchdahl's limit \cite{Buchdahl:1959zz}. This limit establishes the maximum compactness of a star in GR made of a fluid and having an external vacuum Schwarzschild solution, and sets the end of equilibrium configurations and the beginning of gravitational collapse. For every other theory of gravity, including the class of $f(R,T)$ theories considered here, such a limit is expected to be modified.

Unlike in pure GR, the presence of the parameter $\lambda$ in our $f(R,T)$ model makes it harder to find analytical solutions with constant density due to the presence of the pressure term on the right-hand side of (\ref{eq:Mr0}), which prevents its direct integration. Particularizing Eqs.(\ref{eq:Phir0}) and (\ref{eq:Mr0}) to this case ($\lambda\neq 0$ but $\sigma=0$), and rewriting the second in a more convenient form, we get
\begin{eqnarray}
\frac{2A \Phi_r}{r}&=& \kappa^2(1+\lambda)\left(\rho_0+P\right) \label{eq:Phir1} \  \\
\frac{2M_r}{r^2}&=& \kappa^2\left(1+2\lambda\right)\rho_0-\frac{\lambda \kappa^2 (\rho_0+P) }{2}  \label{eq:Mr1} \ .
\end{eqnarray}
We can now use Eq.(\ref{eq:quarkEOS0}) in the form
\begin{equation}\label{eq:quarkEOS1}
\rho_0+P=\frac{\rho_0 A_\odot^{1/\gamma}}{A^{1/\gamma}e^{2\Phi/\gamma}} \ ,
\end{equation}
where now from Eq.(\ref{eq:gamma}) we have $\gamma=(2+3\lambda)/(1+\lambda)$, to rewrite the above equations as 
\begin{eqnarray}
{2A^{\frac{1+\gamma}{\gamma}} e^{2\Phi/\gamma}\Phi_r}&=& \kappa^2(1+\lambda)\rho_0 A_\odot^{1/\gamma} {r}\label{eq:Phir} \  \\
\frac{2M_r}{r^2}&=& \kappa^2\left(1+2\lambda\right)\rho_0-\frac{\lambda \kappa^2 }{2}\frac{\rho_0 A_\odot^{1/\gamma}}{A^{1/\gamma}e^{2\Phi/\gamma}}  \label{eq:Mr} 
\end{eqnarray}
In this form, it is evident that we are now dealing with a non-linear set of first-order coupled differential equations. Interestingly, if we set $\gamma=-1$, the first of the above equations decouples and, at the same time, the second becomes linear in the variable $M(r)$, which does allow us to obtain analytical solutions. This choice of $\gamma$ corresponds to $\lambda=-3/4$, and turns the above equations into 
\begin{eqnarray}
{2 e^{-2\Phi}\Phi_r}&=& \frac{\kappa^2\rho_0}{4 A_\odot} {r}\label{eq:Phir} \  \\
M_r+\frac{3\kappa^2 \rho_0 e^{2\Phi}r}{8A_\odot } M(r)&=& \frac{\kappa^2\rho_0}{4}\left(\frac{3 e^{2\Phi}}{4A_\odot}  -1\right)r^2\ .
\end{eqnarray}
By direct integration of the first, we obtain 
\begin{equation}
e^{-2\Phi(r)}=\left\{\begin{array}{ll}1+\frac{\kappa^2\rho_0}{8A_\odot}(R_\odot^2-r^2)  & r\leq R_\odot \\ 
\hspace{1cm}\\
1 & r> R_\odot \end{array}\right. \ .
\end{equation}
Inserting  this result in the second, we find that 
\begin{eqnarray}\label{eq:Mm3b4}
M(r)&=&(a^2-b^2 r^2)^{3/2}\left[C_1+\frac{\kappa^2\rho_0}{4b^3}\arctan\left(\frac{br}{\sqrt{a^2-b^2 r^2}}\right)\right] \nonumber \\
&-&  \frac{\kappa^2\rho_0 }{4b^2}(a^2-b^2 r^2)r+\frac{\kappa^2\rho_0 r^3}{16 A_\odot a^2} \ ,
\end{eqnarray}
where we have defined $a^2=1+b^2R_\odot^2$ and $b^2=\kappa^2\rho_0/8A_\odot$. The integration constant $C_1$ can be set to zero by noting that $\displaystyle \lim_{r\to 0} M(r)=0$. 

On the other hand, the pressure follows from Eq.(\ref{eq:quarkEOS1}) by setting $\gamma=-1$ and takes the form
\begin{equation}
P(r)=\frac{\rho_0}{A_\odot}\left(A(r)e^{2\Phi(r)}-A_\odot\right) \ .
\end{equation}
Exploring the limit as $r\to 0$ we find that
\begin{equation}
P_c\equiv \lim_{r\to 0} P(r)= \frac{\rho_0}{A_\odot}\left(\frac{1}{1+(b R_\odot)^2}-A_\odot\right) \ .
\end{equation}
To guarantee that $P_c>0$, we must have that the compactness satisfies $\mathcal{C} \equiv M_\odot/R_\odot>\kappa^2\rho_0R_\odot^2/16$, which establishes a constraint between the compactness, the energy density, and the radius of the star. While for constant density stars in GR we found above that $\mathcal{C}_{GR}=\kappa^2\rho_0R_\odot^2/6$, here we see that $\mathcal{C}_{\lambda=-3/4}>\frac{3}{8}\mathcal{C}_{GR}$ for the same values of $\rho_0$ and $R_\odot$. In order to see if there are values of the compactness for which $P_c$ gives divergent results, which happens if and only if the condition 
\begin{equation}
\frac{M_\odot}{R_\odot}=\frac{1}{2}+\frac{\kappa^2\rho_0R_\odot^2}{16}
\end{equation}
is satisfied, we should be able to find an explicit relation  between the compactness $M_\odot/R_\odot $ and the central density $\kappa^2\rho_0$. In the $\lambda=-3/4$ case such a relation can be obtained by evaluating $M(R_\odot)/R_\odot $ using Eq.(\ref{eq:Mm3b4}), which after algebraic manipulations leads to
\begin{equation}
A_\odot +\frac{\kappa^2\rho_0 R_\odot^2}{8}=\frac{1}{\left(\frac{\arctan\left[b R_\odot\right]}{b R_\odot}-\frac{3}{4}\right)} 
\end{equation}
Unlike in GR, here the relation between $M_\odot/R_\odot$ and $\kappa^2\rho_0R_\odot^2$ is highly non-linear, which prevents its explicit analytical resolution. Proceeding perturbatively, assuming that $\kappa^2\rho_0 R_\odot^2$ is small and that $A_\odot$ is of order unity, we can approximate $\arctan\left[b R_\odot\right]/b R_\odot\approx 1$ to obtain 
\begin{equation}
\frac{M_\odot}{R_\odot}=-\frac{3}{2}+\frac{\kappa^2\rho_0 R_\odot^2}{16} \ ,
\end{equation}
which does not make any physical sense because it would imply a negative mass object with a very strong gravitational field despite being in the weak density regime. This unphysical result can be understood from the fact that the value $\lambda=-3/4$ implies that the contribution from the energy density to the mass in Eq.(\ref{eq:Mr1}) is negative. The effect on Eq.(\ref{eq:Phir1}) is not so dramatic because the $(1+\lambda)$ term does not flip its sign for $\lambda>-1$. In any case, all our attempts to find solutions for $\mathcal{C}_{\lambda=-3/4}$ as a function of $\kappa^2 \rho_0 R_\odot^2$ lead to complex solutions, which confirms the lack of physical reality of this constant-density object.

\section{Quark matter stars}\label{sec:Quark}
We will now consider the case with $\alpha=1$ and $\sigma\neq 0$ in Eqs.(\ref{eq:Phir}) and (\ref{eq:Mr}) (related to the equation of state of quark matter \cite{Mak:2003kw}), which turns them into 
\begin{eqnarray}
 \frac{2M_r}{r^2}&=& \kappa^2\left(1+\frac{3\lambda}{2}\right)\rho_0+\frac{\kappa^2 P }{2} \left((2+3\lambda )\sigma-\lambda \right) \label{eq:MrS} \  \\
\frac{2A \Phi_r}{r}&=& \kappa^2(1+\lambda)\left(\rho_0+(1+\sigma)P\right) \label{eq:PhirS}
 \ .
\end{eqnarray}
The freedom introduced by the parameter $\sigma$ allows us to cancel the pressure-dependent term of the mass equation by setting $\sigma=\lambda/(2+3\lambda)$, and the above equations become
\begin{eqnarray}
 \frac{2M_r}{r^2}&=& \kappa^2\left(1+\frac{3\lambda}{2}\right)\rho_0 \label{eq:MrS1} \  \\
\frac{2A \Phi_r}{r}&=& (1+\lambda)\kappa^2\rho_0\frac{A_\odot^{1/2}}{A^{1/2}e^\Phi} \label{eq:PhirS}
 \ ,
\end{eqnarray}
where we have used the relation 
\begin{equation}\label{eq:quarkEOS}
\rho_0+(1+\sigma)P=\rho_0 \frac{A_\odot^{1/2}}{A^{1/2}e^{\Phi}} \ ,
\end{equation}
as well as the fact that for this value of $\sigma$ we have $\gamma=2$, like in the constant-density stars of GR. Note that this value of $\sigma$ allows us to consider all possible values of $\lambda$, which gives us hopes to find physically meaningful solutions, at least in the limit of small $\lambda$.

From the above equations we find 
\begin{eqnarray}
M(r)&=& \frac{\kappa^2\rho_0(2+3\lambda)}{12}r^3 =\frac{M_\odot r^3}{R_\odot^3}\ \\
e^{\Phi(r)}&=& \frac{3(1+\lambda)}{2+3\lambda}\left[\frac{A_\odot^{1/2}}{A(r)^{1/2}}-\frac{1}{3(1+\lambda)}\right] \ ,
\end{eqnarray}
where we see that the effective Schwarzschild mass $M_\odot$ is positive for any $\lambda>-2/3$. The line element associated with these solutions takes the form
\begin{equation}
ds^2=\frac{\left(3(1+\lambda)A^{1/2}_\odot-A^{1/2}(r)\right)^2}{(2+3\lambda)^2}dt^2-\frac{1}{A(r)}dr^2-r^2d\Omega^2 \ ,
\end{equation}
where $A(r)$ is defined as in Eq.(\ref{eq:AGR}) but with the identification $M_\odot/R_\odot^3\equiv {\kappa^2\rho_0(2+3\lambda)}/{12}$.

\begin{figure}[t!]
\includegraphics[width=8.4cm,height=5.5cm]{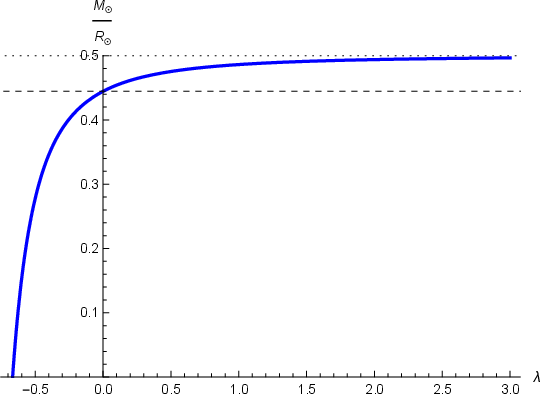}
\caption{Compactness $M_{\odot}/R_{\odot}$ for a quark matter star in $f(R,T)=R+\kappa^2 \lambda T$ gravity theory as a function of the $\lambda$ parameter. The dashed line denotes the compactness (Buchdahl) limit for the GR case, $M_{\odot}/R_{\odot}=4/9$, while the dotted line corresponds to $M_{\odot}/R_{\odot}=1/2$, which is the compactness of a black hole. As we can see, for $\lambda>0$ we can have stars more compact than the most compact stars of GR but less compact than a Schwarzschild black hole.}
\label{fig:com}
\end{figure}

The pressure can be obtained from Eq.(\ref{eq:quarkEOS}) particularized to $\sigma=\lambda/(2+3\lambda)$ and leads to
\begin{equation}\label{eq:quarkEOS2}
P(r)=\frac{(2+3\lambda)\rho_0}{2(1+2\lambda)} \left[\frac{A(r)^{1/2}-A_\odot^{1/2}}{3(1+\lambda)A_\odot^{1/2}-A(r)^{1/2}}\right] \ ,
\end{equation}
The maximum central density occurs when the denominator vanishes, and this implies that $3(1+\lambda)A_\odot^{1/2}=1$. As a result, the critical compactness limit for this family of stars is given by 
\begin{equation}
\frac{M_\odot}{R_\odot}=\frac{1}{2}\left(1-\frac{1}{9(1+\lambda)^2}\right) \ .
\end{equation}
In the limit $\lambda\to 0$, the above expression recovers the Buchdahl limit of GR, namely, $M_\odot/R_\odot=4/9$. As depicted in Fig. \ref{fig:com}, in the $f(R,T)$ case, we see that for $\lambda>0$ all solutions surpass the GR compactness limit, which implies that this kind of modified gravity can accommodate more compact stars than in GR if $\lambda>0$. This is true even in the limit $\lambda \to \infty$, in which the compactness of these stars asymptote to the one of a black hole, namely, $M_\odot/R_\odot=1/2$. Conversely, for $\lambda <0$ the compactness is smaller than in the GR case, saturating at vanishing compactness for $\lambda=-2/3$. In any case, from a physical point of view the size of $\lambda$ must be small, as it can be constrained using e.g. studies of geodesic deviation.

\section{Final remarks} \label{sec:Final}

In this work, we have considered the existence of analytical solutions of static, spherically symmetric stellar objects in a particularly simple $f(R,T)$ theory of gravity, namely, $f(R,T)=R+\lambda \kappa^2 T$. Following the literature, we have first considered the case of constant-density stars, identifying a particular value of $\lambda$ for which the relevant equations can be decoupled and linearized. Unfortunately, this case leads to some inconsistencies that make such solutions unacceptable from a physical perspective. We then moved to a family of solutions characterized by a quark matter equation of state, in which the density is no longer constant but has a linear dependence with the fluid pressure. In this case, one can choose the proportionality constant of the pressure term to decouple the equations and get analytical solutions. The result makes much more physical sense and represents a simple deformation of the constant-density solutions of GR. It also implies that more compact stars than in GR are possible for standard quark matter equations of state if new gravitational dynamics of the $f(R,T)$ type is considered. Whether this could lead to observational discriminators between these families of theories and GR is yet to be investigated.

These analytical solutions provide complementary information to the bunch of numerical analyses made in the literature of the subject \cite{Olmo:2019flu}, particularly on the compactness of the corresponding stars, using the minimal extension of GR of this class. Such an extension can be further generalized to more involved models such as $f(R,T)=R+\lambda (\kappa^2T)^n$, with $n>1$ some integer number. From our results it should now be feasible the analysis of limiting masses within these theories, such as the minimum hydrogen burning mass (as already done on its $f(R)$ counterpart, see e.g. \cite{Olmo:2019qsj,Olmo:2021yul}), on which we hope to report soon. 

\section*{Acknowledgments}

This work is supported by the Spanish National Grants  PID2020-116567GB-C21 and PID2022-138607NBI00, funded by MCIN/AEI/10.13039/501100011033 (``ERDF A way of making Europe" and ``PGC Generaci\'on de Conocimiento"). 
This article is based upon work from COST Action CA21136, funded by COST (European Cooperation in Science and Technology).

\bibliographystyle{apsrev4-1}

\end{document}